# On the Secrecy Rate of Interference Networks Using Structured Codes

Shweta Agrawal
University of Texas, Austin
ECE Department
Austin, TX, USA
Email: sagrawal@ece.utexas.edu

Sriram Vishwanath
University of Texas, Austin
ECE Department
Austin, TX, USA
Email: sriram@ece.utexas.edu

*Abstract*— This paper shows that structured transmission schemes are a good choice for secret communication over interference networks with an eavesdropper. Structured transmission is shown to exploit channel asymmetries and thus perform better than randomly generated codebooks for such channels. For a class of interference channels, we show that an equivocation sum-rate that is within two bits of the maximum possible legitimate communication sum-rate is achievable using lattice codes.[1]

## I. INTRODUCTION

Secret communication over networks is an increasingly important field of research, with multiple (information theoretic, networking and cryptographic) perspectives on this problem. From an information theoretic perspective, the capacity of a wiretap channel, a model for eavesdropping attack on a point to point channel, was introduced and analyzed in [7], [6]. A combination of random-coding[2] with binning was used to achieve the capacity of the wiretap channel. Subsequently, structured binning arguments (as in [3]) were used instead of random binning to obtain the same results for wiretap channel capacity.

Subsequently, the wiretap channel framework has been generalized to multiple other settings, including channels such as the multiple access [4] and cognitive-interference [5] channels and also to network models such as the wireless erasure network [8]. In all of these cases, as in [6], random coding and binning arguments form the basis for achieving the secrecy rate (region).

In recent years, there is an increasing interest in structured coding schemes, particularly for enhancing the rates achievable in interference/wireless networks [13], [10]. Lattice codes are shown to induce an *alignment* in the interference seen at each node, thus enhancing the degrees of freedom [10] and the secure degrees of freedom for an interference network [9]. Simultaneously, lattice coding has also been shown to possess security benefits, for information hiding [12] and for information relaying in wireless networks [1], [2].

In this work, we emphasize the gains of using lattice coding schemes over random coding schemes for the interference

[1]This work is supported by DARPA IAMANET and a grant from the Air Force Office of Sponsored Research

[2]where a codebook is chosen from a family of codebooks generated using i.i.d. realizations of a suitable random variable

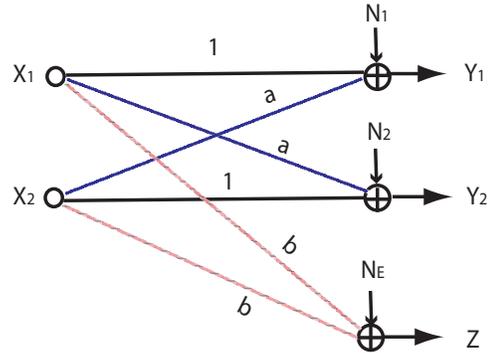

Fig. 1. The system model with two legitimate transmit receive pairs and one eavesdropper. The legitimate receivers observe unequal signal strengths from the two transmitters, while the eavesdropper observes equal channel gains.

channel with an eavesdropper. Specifically, we show that, for a class of interference channels, lattice codes achieve an equivocation rate for each user that is within one-bit of its individual rates. In other words, the eavesdropper gains no more than one bit of information about the legitimate messages per transmission. We also argue that randomly generated codebooks, in general, cannot achieve this equivocation rate. The rest of this paper is organized as follows. In section II we motivate and present the system model. In Section III, we present the main results for 1 bit secrecy. In section V, we discuss the advantages of using structured coding versus random coding from the secrecy perspective. We conclude with Section VI.

## II. SYSTEM MODEL

The notation used in this paper is as follows. $|.|$ denotes the cardinality of a set, $\oplus$ is used to denote the direct/Minkowski sum of two sets, i.e.

$$S_1 \oplus S_2 \triangleq \{s_1 + s_2, s_1 \in S_1, s_2 \in S_2\}.$$

$\mathbb{Z}^n$ denotes the integer lattice in $n$ dimensions, and $\mathbb{Z}_p^k$ denotes the set of all $k$ length vectors from $GF(p)$.

The system model is shown in Figure 1. In this setting, we consider a symmetric interference channel as the legitimate

communication network, and an external eavesdropper. The legitimate channel is mathematically given by:

$$Y_1 = X_1 + aX_2 + N_1 \\ Y_2 = X_2 + aX_1 + N_2 \quad (1)$$

where $a \neq 1$. The eavesdropper's channel is given by:

$$Z = bX_1 + bX_2 + N_e \quad (2)$$

Here, $X_i, i \in \{1,2\}$, is the output signal from Transmitter $i$, which is assumed to be constrained to an average power of $P$. Legitimate Receiver $i$ observes $Y_i, i \in \{1,2\}$, and is assumed to suffer from additive Gaussian noise of unit variance. The eavesdropper receives $Z$, which is an equally weighted linear combination of the two transmit signals with additive Gaussian noise of variance $N_e$. In this paper, we make no assumptions on the value of $N_e$ or the joint distribution between the noise variables $N_1, N_2$ and $N_e$. In effect, the secrecy results presented in Section III hold even when $N_e = 0$.

Note that the legitimate channels between the transmitters and the legitimate receivers ($p(y_i|x_1,x_2), i \in \{1,2\}$) are *different* from the channel between the transmitters and the eavesdropper ($p(z|x_1,x_2)$). This asymmetry in channels is exploited by the structured coding strategy developed in Section III. In other words, if $a = 1$, then lattice coding strategy described in Section III does not, in general, provide an advantage over randomly generated codebooks in enabling secure communication.

The goal is to communicate at the highest sum-rate possible in the legitimate channel while ensuring that the eavesdropper's understanding of the legitimate messages is constrained. If $W_1, W_2$ represent the legitimate messages with cardinalities $2^{nR_1}$ and $2^{nR_2}$ respectively, then we desire to

$$\max(R_1 + R_2) \quad (3)$$

such that $(R_1, R_2)$ belong to the achievable region of the legitimate interference channel. Let $R_{ei}$ denote the equivocation rate for each message:

$$R_{ei} = \frac{1}{n} H(W_i | Z^n)$$

for $i \in \{1,2\}$. The 1 bit secrecy requirement can be defined as $R_{ei} \geq R_i - 1$ for $i \in \{1,2\}$. This is equivalent to $\frac{1}{n} I(W_i; Z^n) \leq 1$, and this is the formulation we will use in the paper. In this paper, we focus on achieving a symmetric rate point of the region defined by (3). Given the symmetry in the channel definition, a symmetric rate point exists that achieves the sum-capacity of this channel, and our goal is to use lattice codes to achieve this rate point for a class of symmetric interference channels.

### III. MAIN RESULTS

First, we require a few lemmas:

*Lemma 1:* Let $C = \Lambda \cap \mathcal{V}_L$ where $\Lambda$ is a "good" Construction-A lattice [3] of dimension $n$ and $\mathcal{V}_L$ is the Voronoi region of a "good" nested lattice $L$ [11]. Let $X_1$ and $X_2$ be random variables distributed uniformly over $C$. Then,

$$H(X_1 + X_2) \leq \log |C| + n$$

[3] By "good", we mean a lattice that is simultaneously good for source and channel coding, as defined in [11].

Proof: Let $X$ be a random variable distributed uniformly over $C \oplus C$. Since $H(X_1 + X_2) \leq H(X) = \log |C \oplus C|$ it suffices to show that

$$|C \oplus C| \leq 2^n |C| \quad (4)$$

The proof of this is comes from the construction of "good" nested lattice codebooks in [11]. Let

$$\Lambda = G'(p^{-1} G \mathbb{Z}_p^k + \mathbb{Z}^n) \\ L = G' \mathbb{Z}^n$$

where $G'$ is an invertible $n \times n$ matrix and $G$ is an $n \times k$ matrix with elements from GF($p$). The details behind this construction and the fact that matrices $G, G'$ exist such that these lattices are "good" can be found in [11], [18].

Then we have,

$$\Lambda \mod L \equiv G'[p^{-1} G \mathbb{Z}_p^k + \mathbb{Z}^n] \mod G'(\mathbb{Z}^n)$$

where modulo $\mathbb{Z}^n$ is intersection with the Voronoi region of $\mathbb{Z}^n$ (as defined in [11]), and $\equiv$ denotes the fact that there is an invertible transformation between the two sets.

Let

$$\gamma \in \Gamma \triangleq [p^{-1} G \mathbb{Z}_p^k + \mathbb{Z}^n] \mod \mathbb{Z}^n$$

Note that the Voronoi region of $\mathbb{Z}^n$, which we denote by $\Omega_{\mathbb{Z}^n}$, is an $n$-dimensional cube centred around the origin, extending between $[-0.5, 0.5]$ in each dimension. Hence by intersecting $\Gamma$ with $\Omega_{\mathbb{Z}^n}$, gives that each (scalar) element of $\gamma$ is contained within the interval $[-0.5, 0.5]$. This implies that every element $\beta \in \Gamma \oplus \Gamma$ is such that $\beta$'s component along each dimension is within the interval $[-1, 1)$. Thus we have:

$$\Gamma \oplus \Gamma \subset [p^{-1} G \mathbb{Z}_p^k + \mathbb{Z}^n] \mod 2\mathbb{Z}^n$$

or that

$$C \oplus C \subset [G'(p^{-1} G \mathbb{Z}_p^k + \mathbb{Z}^n)] \mod 2L$$

which, in turn, establishes (4) and thus the lemma.

*Lemma 2:* Let $X_1, X_2$ be two independent random variables uniformly distributed over $C$. Then,

$$\frac{1}{n} I(X_1; X_1 + X_2) \leq 1$$

Proof: Note that

$$H(X_1 | X_1 + X_2) = H(X_1, X_2) - H(X_1 + X_2) \\ = H(X_1) + H(X_2) - H(X_1 + X_2) \\ \geq 2 \log |C| - (\log(|C|) + n) \\ \geq \log |C| - n$$

Hence,

$$I(X_1; X_1 + X_2) = H(X_1) - H(X_1 | X_1 + X_2) \\ \leq \log |C| - \log |C| + n$$

## A. Very Strong Interference Channel

The "very strong" interference channel was first studied in [14]. By very strong, we mean that the interference channel gain ($a$ in equation (1)) is large enough such that the interfering signal (for e.g. from Transmitter 1) can be successfully decoded at the receiver (Receiver 2), thus resulting in parallel channels. For the class of very strong interference channels for which $a^2 \geq P+1$, we show that rates arbitrarily close to the the unconstrained (without the secrecy requirement) capacity region of the channel can be achieved using lattice codes with 1-bit secrecy for the channel given by Figure 1.

Symmetric interference channels with multiple ($> 2$) transmit-receive pairs and very strong interference are investigated in [15] using a lattice coding framework. In [15], it is shown that if $a^2 \geq \frac{(P+1)^2}{P}$, then codebooks that are nested subsets of the same lattice can be used such that each legitimate communication channel achieves its single user rate. For the case of the symmetric rate point, these codebooks become *identical*. Thus, we have:

$$
\begin{aligned}
I(W_i; Z^n) &\overset{(a)}{\leq} I(L_i; Z^n) \\
&\overset{(b)}{\leq} I(L_i; L_1 + L_2) \\
&\overset{(c)}{\leq} n
\end{aligned}
\quad (5)
$$

where $L_1, L_2$ are uniformly distributed random variables over the common lattice-based codebook used at each transmitter. Also, $(a)$ and $(b)$ are due to the data processing inequality and $(c)$ is due to Lemma 2.

## B. Weak Interference Channels

Weak interference channels, where $a < 1$, have received significant attention in recent years. In [16], the authors determine that, for all $a$ such that $|a + a^3 P| \leq \frac{1}{2}$, treating interference as noise in the legitimate (unconstrained) channel is optimal, i.e., that

$$R_1 = R_2 = \frac{1}{2} \log \left(1 + \frac{P}{a^2 P + 1}\right)$$

is achievable. The achievability argument in [16] uses random coding. Here, we show that (nested) identical lattice codebooks can be used to achieve the same set of rates, and thus, using identical steps as those in (5), we can establish that the unconstrained sum-capacity can be achieved in this channel even with a 1-bit secrecy constraint.

Consider "good" nested lattices $\Lambda_c \subset \Lambda_f$ (as defined in [11]), and define the codebook

$$C \triangleq \Lambda_f \cap \mathcal{V}_{\Lambda_c}$$

where $\mathcal{V}_{\Lambda_c}$ is the Voronoi region of the lattice $\Lambda_c$. Here, $\Lambda_c$ represents the power constraint on each transmitter and $\Lambda_f$ is chosen to enable successful decoding in the presence of noise at the receiver. Given $L_i \in C$ at Transmitter $i$, $i \in \{1, 2\}$, we construct

$$X_i = [L_i + U_i] \mod \Lambda_c,$$

where $U_1, U_2$ are two independent random vectors chosen uniformly over $\mathcal{V}_{\Lambda_c}$, and the mod operation is as defined in [11].

At each receiver, we determine

$$Y_i' = [\alpha Y_i - U_i] \mod \Lambda_c$$

where

$$\alpha = \frac{P}{(1+a^2)P + N}$$

At Receiver 1, following the analysis in [11], we have

$$
\begin{aligned}
Y_1 &= [\alpha Y_1 - U_1] \mod \Lambda_c \\
&= [\alpha Y_1 - L_1 - U_1 + L_1] \mod \Lambda_c \\
&= [\alpha Y_1 - [L_1 + U_1] \mod \Lambda_c + L_1] \mod \Lambda_c \\
&= [L_1 + \alpha Y_1 - X_1] \mod \Lambda_c \\
&= [L_1 + (1-\alpha)X_1 + \alpha(X_2 + N_1)] \mod \Lambda_c \\
&= [L_1 + N_1'] \mod \Lambda_c
\end{aligned}
$$

where

$$N_1' \triangleq (1-\alpha X_1) + \alpha(X_2 + N_1)$$

is the effective noise with variance

$$\frac{P(a^2 P + N)}{(1+a^2)P + N}$$

that the receiver which is independent of $L_1$. Based on the analysis in [18], [11], the lattice point $L_1$ can be determined with high reliability if

$$R_1 < \frac{1}{2} \log \left(1 + \frac{P}{a^2 P + 1}\right).$$

A similar argument can be used to analyze the decoding at Receiver 2 to achieve the same rate. In effect, for this setting (and a few other multiple access/broadcast/interference channel settings) random coding arguments for Gaussian channels that use randomly generated codebooks for encoding followed by successive decoding at the receiver can be replaced with lattice coding and decoding arguments to obtain the same rates. Since in the above coding strategy, the *same* codebook $C$ is used at each transmitter, we can follow the same steps as in 5 to get the desired 1-bit secrecy: $I(W_i; Z^n) \leq n$.

## C. General Interference Channels

Here we consider more general Gaussian interference channel settings, i.e. for values of $a$ that are neither very strong nor very weak. For an interference channel with two transmit-receive pairs, the nested lattice coding strategy described in [10] achieves a rate region that is, in general, a subset of the Han-Kobayashi region [19]. Thus, for general values of $a$, it is not yet known if lattice codebooks can be used to achieve the same rate region for the unconstrained two-user Gaussian interference channel as the Han-Kobayashi region. However, we show that even if lattice codebooks are not necessarily capacity achieving, they provide the same secrecy benefits for interference channels with a wiretapper (when $a \neq 1$) as the cases analyzed in subsections III-A and III-B.

We now describe the lattice coding scheme and investigate its secrecy capacity. As depicted in Figure 1, Transmitter $T_j$ where $j \in \{1, 2\}$, wants to send a message $m_j \in \{1, \ldots, 2^{nR}\}$ to receiver $R_j$. As in [10], Transmitter $T_j$, divides its message $m_j$ into $N$ parts $m_{j1}, \ldots, m_{jN}$ so that rate $R_i$ is associated with the $i^{th}$ message part $m_{ji}$. Each of the $N$ submessages $m_{ji}$ is encoded to $X_{ji}^n$ using a distinct lattice codebook $C_i$. Note that the set of $N$ lattice codebooks used by each transmitter say $C_1, \ldots, C_N$, is *identical*. Both transmitters assign power $P_i$ to $m_{ji}$. Transmitter $T_j$ transmits $m_j$ as $\sum_{i=1}^N X_{ji}^n$. Without loss of generality (due to symmetry), we focus on the first transmitter-receiver pair.

The receiver $R_1$ ($R_2$ is handled similarly) receives

$$Y_1^n = \sum_{i=1}^N X_{1i}^n + \sum_{i=1}^N X_{2i}^n + N_1$$

where $N_1$ is the noise. $R_1$ successively decodes each of the $N$ submessages to recover the whole message $m_j$. By choosing power assignment $P_i$ cleverly depending on the value of $a$ (see [10] for details), the "very strong" interference condition can be satisfied at each of the $N$ stages, so that for each stage the interference can be decoded and subtracted, then the submessage can be recovered. It is shown in [10], that by careful selection of lattice codebooks for each stage, the message $m_1$ can be recovered from received codeword $\sum_{i=1}^N X_{1i}^n$, with vanishingly small probability of error (vanishes as $n^{\frac{-n}{2}}$).

Thus, for the rate region described in [10], as lattice dimension $n \to \infty$, the tuple of $N$ lattice points ($i^{th}$ point corresponding to the $i^{th}$ submessage), can be recovered with arbitrarily small probability of error from the sum of these points. This implies that for large lattice dimensions, there is "almost" a one-one mapping between an $N$-tuple of lattice points (chosen from the $N$ special lattices of our coding scheme) and the componentwise sum of these $N$ points. Note that since the probability of error is nonzero, it is possible that two $N$-tuples sum to the same value, but there can be only very few of such points, and they are negligible as the dimension of the lattice goes to infinity. This leads to the following lemma:

*Lemma 3:* Let $X_i$ be picked uniformly at random from codebook $C_i$, for $(i = 1, 2, \ldots, N)$, then the distribution of $X = \sum_{i=1}^N X_i$ converges as $n \to \infty$ (in the weak sense) to a uniform distribution over $C = C_1 \oplus C_2 \oplus \ldots \oplus C_N$.

To show one bit secrecy, we next need the following lemma:

*Lemma 4:* Let $L_1$ and $L_2$ be two random variables distributed as $X$. Then,

$$H(L_1 + L_2) \leq \log |C| + n$$

Proof: By lemma 3, we have that $L_1$ and $L_2$ are distributed asymptotically uniformly on $C$. Also, the codebook $C = C_1 \oplus C_2 \oplus \ldots \oplus C_N$, due to nesting, lies in the same lattice $\Lambda$ of lemma 1 which is a "good" Construction A lattice of dimension $n$. The proof of the claim thus follows from the proof of Lemma 1, and is omitted here for brevity.
Thus, by Lemmata 2, 3 and 4, we have that

$$\frac{1}{n} I(L_1; L_1 + L_2) \leq 1$$

Now, we use steps identical to 5 to obtain the desired 1-bit secrecy as $\frac{1}{n} I(W_i; Z^n) \leq 1$.

## IV. SIMPLER PROOF FOR 1-BIT SECRECY

In this section, we provide another proof that 1-bit secrecy can be achieved by lattice codes.
*Theorem 1:*
$$I(W_1|Z^n) \leq n$$

*Proof:* Consider that each transmitter has a lattice codebook $C$ which contains $2^{nR}$ lattice points. Place them uniformly in $2^{n\hat{R}}$ bins, where $\hat{R} < R$. The message to be transmitted is the bin index, and the codeword for the message $w$ is a lattice point picked uniformly from bin $w$.

The eavesdropper sees

$$Z^n = b(X_1^n + X_2^n) + N_e^n$$

but by DPI,

$$I(W_1; Z^n) \leq I(W_1, \hat{Z}^n)$$

where $\hat{Z}^n = X_1^n + X_2^n$.
Thus, it suffices to show that $I(W_1, \hat{Z}^n) < n\epsilon$. We have:

$$
\begin{aligned}
H(W_1|\hat{Z}^n) &= H(W_1|\hat{Z}^n) - H(W_1|\hat{Z}^n, X_1^n) \\
&= I(W_1; X_1^n|\hat{Z}^n) \\
&= H(X_1^n|\hat{Z}^n) - H(X_1^n|W_1, \hat{Z}^n) \\
&\geq n(R-1) - H(X_1^n|W_1, \hat{Z}^n) &(6) \\
&\geq n(R-1) - H(X_1^n|W_1) &(7) \\
&\geq n(R-1) - \{H(W_1|X_1^n) + H(X_1^n) - H(W_1)\} &(8) \\
&\geq n(R-1) - \{0 + nR - n\hat{R}\} &(9) \\
&\geq n(R - 1 - R + \hat{R}) \\
&\geq n(\hat{R} - 1) &(10) \\
& &(11)
\end{aligned}
$$

$$
\begin{aligned}
I(W_1; Z^n) &= H(W_1) - H(W_1|Z^n) &(12) \\
&= n\hat{R} - H(W_1|Z^n) &(13) \\
&\leq n\hat{R} - n(\hat{R} - 1) &(14) \\
&= n &(15)
\end{aligned}
$$

as desired. ∎

## V. THE NEED FOR STRUCTURED ENSEMBLES OVER RANDOMLY GENERATED ENSEMBLES

Notice that in the coding arguments presented above, the codebooks utilized at each of the transmitters are identical. This coupling, along with the linear structure, are important features of the lattice coding argument. To explain this further, let us reconsider the symmetric interference channel with a wiretapper (Figure 1) and use randomly generated codebooks as defined in [17] for the legitimate channel.

Note that randomly generated codebooks with joint (typical-set) decoding is the conventional achievability argument for the multiple-access channel. As shown in [17], as long as:

$$R_1 + R_2 \leq \frac{1}{2} \log \left( 1 + \frac{b^2(P_1 + P_2)}{N_e} \right)$$

along with individual rate bounds on each $R_1$ as given in [17], the probability of error, averaged over all randomly generated codebooks, can be made arbitrarily small. This clearly does not mean that each codebook has a small probability of error (averaged over all codewords). It does imply that a large fraction of these codebooks are good choices for this channel. In essence, as the block length of the codebook is increased, a larger fraction of the set of all randomly generated codebooks can be chosen to meet the probability of error constraint.

This is the main reason that randomly generated codebooks are, in general, bad choices for communicating over the interference channel with a wiretapper. Although there might exist a particular random generated codebook that is "good" for the legitimate channel and "bad" for the eavesdropper, for sufficiently large $b$ (or correspondingly, sufficiently small $N_e$), most of the randomly generated codebooks provide limited to no secrecy to the legitimate interference channel. As noted in the previous sections, the 1-bit secrecy afforded by lattice codes holds regardless of the value of $b$ or of $N_e$.

**A Note on Computational Secrecy:** Note that there is a computational aspect to this secrecy problem (a.k.a. weak secrecy). As established in [20], low density lattice codes exist that can be encoded and decoded in polynomial time. As the channel codes considered here result in *nested* lattices at the receiver, lattice decoding is sufficient for determining the lattice point and thus the message being communicated, and can thus be implemented in polynomial-time.

The eavesdropper, on the other hand, can at best determine the sum of the two lattice points being communicated. As there exist an exponential number of possible lattice pairs that sum to the same value, even if the eavesdropper successfully determines the sum-lattice point, it is exponentially hard for it to enumerate (and therefore, even locate) the pair of lattice points being communicated in the legitimate system.

## VI. CONCLUSIONS

In this paper, we consider an interference channel with a wiretapper. The key feature of this model is that the channel between the transmitters and legitimate receivers and the channel between the transmitter and the eavesdropper are different. The fact that channel gain $a \neq 1$ leads to nested lattices at the receiver, which can be decoded, however the eavesdropper sees a sum of 2 distinct (not nested) lattices, which do not reveal information. Thus, this asymmetry is exploited by structured coding schemes, while random coding is, in general, blind to it.

As a next step, we plan to investigate a larger class of multi-source networks, and utilize channel asymmetries, along with structured codes, to enhance the rate at which secure communication is possible over these networks.